# Hierarchical and Nonhierarchical Three-Dimensional Underwater Wireless Sensor Networks

S. M. Nazrul Alam
Department of Computer Science
Cornell University, Ithaca, NY, USA
Email: S.M.Nazrul.Alam@cornell.edu

Zygmunt J. Haas
School of Electrical and Computer Engineering
Cornell University, Ithaca, NY, USA
Email: haas@ece.cornell.edu

**Abstract:** In some underwater sensor networks, sensor nodes may be deployed at various depths of an ocean making those networks three-dimensional (3D). While most terrestrial sensor networks can usually be modeled as two dimensional (2D) networks, these underwater sensor networks must be modeled as 3D networks. This leads to new research challenges in the area of network architecture and topology. In this paper, we present two different network architectures for 3D underwater sensor networks. The first one is a hierarchical architecture that uses a relatively small number of robust backbone nodes to create the network where a large number of inexpensive sensors communicate with their nearest backbone nodes, and packets from a backbone node to the sink is routed through other backbone nodes. This hierarchical approach allows creating a network of smaller number of expensive backbone nodes while keeping the mobile sensors simple and inexpensive. Along with network topology, we also study energy efficiency and frequency reuse issues for such 3D networks. The second approach is a nonhierarchical architecture which assumes that all nodes are identical and randomly deployed. It partitions the whole 3D network space into identical cells and keeps one node active in each cell such that sensing coverage and connectivity are maintained while limiting the energy consumed. We also study closeness to optimality of our proposed scheme.

## 1. INTRODUCTION

Two major differences between terrestrial sensor networks and underwater sensor networks are network dimensionality (i.e., 3D instead of 2D) and communication medium (i.e., acoustic instead of radio). Terrestrial wireless sensor networks, where sensors are deployed on earth surface and where the height of the network is smaller than the transmission radius of a node, can usually be modeled as two-dimensional (2D) networks. However, in many underwater sensor networks, nodes may be placed at different depths of an ocean and thus these networks must be modeled as three-dimensional (3D) networks [1]. In many cases, network architecture and topology directly depend on the physical dimensionality of the network and so results from 2D terrestrial sensor networks can not be used in 3D underwater sensor networks. In this paper, we try to address this issue and propose two approaches to build a 3D underwater sensor network. The first one is a hierarchical network architecture where the network is maintained by a small number of robust and powerful backbone nodes. Actual sensing is done by large number of inexpensive and failure prone sensor nodes. Once the sensing information reaches any backbone node, the information is routed to the sink through other backbone nodes. Then we investigate frequency reuse issues for such 3D networks including variations due to acoustic communication. We also study energy efficiency of different hierarchical network topologies. The second approach creates a nonhierarchical network using one type of nodes. This scenario is applicable where a large number of nodes are randomly and uniformly deployed in the network space. We exploit redundancy to improve network lifetime by partitioning the 3D network space into identical cells such that only one node remains active in each cell and full coverage and connectivity is maintained. We then extend our work for *k*-coverage, where any point in the network has to be within the sensing range of at least *k*-nodes. We also provide comparison between our proposed scheme and the scheme where an oracle can decide where to place the nodes (i.e., the *optimal scheme*). Our scheme has better performance for higher values of *k*. For example, our study shows that our scheme can provide 4-coverage with probability 0.9971 with twice the nodes needed by the optimal scheme.

The rest of the paper is organized as follows. Related works are described in Section 2. Section 3 presents the proposed hierarchical network architecture as well as frequency reuse issues and energy efficiency issues for 3D networks. Section 4 describes the proposed nonhierarchical network approach and shows its closeness to optimality. Discussions on routing issues and possible future research directions are described in Section 5. The paper is concluded in Section 6.

## 2. RELATED WORKS

Recent interest in three-dimensional underwater sensor networks have generated a lot of research endeavors in 3D networking [20][21][8][19][9][11].

Coverage and connectivity issues of 3D networks have been explored in [2] and the solution provided in that paper works when the ratio of communication and sensing range is at least 1.7889. This result has been generalized for any ratio of communication and sensing range in [3]. In this paper, we use these results as a basis to determine where to place backbone nodes in our hierarchical network.

Frequency reuse for underwater acoustic sensor networks have been investigated in [23][24]. However, that work assumes that the network is two-dimensional. In this paper, we extend that work for a 3D scenario. Frequency planning in 3D for radio network has been investigated in [10].

Since some of the the analysis in this paper uses the concepts of polyhedron, space-filling polyhedron, Kelvin's conjecture, and Voronoi tessellation, now we provide very brief references on them. Details on these concepts are available in [2]. A polyhedron is a three-dimensional shape that consists of finite number of polygonal faces. A space-filling polyhedron is a polyhedron that tessellates a 3D space. It is hard to show that a polyhedron has space-filling property. For example, Aristotle himself made a mistake when claimed that the tetrahedron fills space [4] and that mistake remained unnoticed until the 16th century [14][17]. Cube is the only space-filling regular polyhedron [12]. Triangular prism, hexagonal prism, cube, truncated octahedron [22][27], and gyrobifastigium [15] are the only five convex polyhedrons with regular faces that have space-filling property. The rhombic dodecahedron, elongated dodecahedron, and squashed dodecahedron are also space-fillers.

Kelvin's conjecture [25] has been used in [2] to show that if 3D network space is divided into truncated octahedron cells such that the radius of each cell is equal to the sensing range of each node, and a node is placed at the center of each cell, then this placement strategy requires minimum number of nodes. Kelvin's conjecture was generally accepted a true for than a century [28], but a counter example was found in 1993 when it is shown that a space-filling structure consisting of six 14-sided polyhedrons and two 12-sided polyhedrons with irregular faces of equal volume has 0.3% less surface area than truncated octahedron [26]. But it is still unknown if Kelvin's conjecture is true for identical cells. Anyway, in a different context it has been shown that body-centered cubic (BCC) lattice has the smallest mean squared error of any lattice quantizer in three dimensions [5], which validates the results of [2].

The aim of our nonhierarchical network architecture is mainly to conserve energy by keeping a subset of the nodes active in a dense network to perform all necessary functions while putting the rest of the nodes into sleep. This is a very common approach in terrestrial sensor networks [29][7][31][30][6]. One important work in this context is geographic adaptive fidelity (GAP) [29] which only works for 2D network. Our nonhierarchical architecture solves that problem for 3D networks. It has been shown that GAP is quite inefficient in terms of number of nodes being kept active. However, in this paper we show that the efficiency of our 3D solution is higher than the efficiency of GAP in 2D.

## 3. HIERARCHICAL NETWORK ARCHITECTURE

In this section, we describe a hierarchical 3D network architecture that has two different types of nodes – more powerful and robust backbone nodes, and less powerful and failure prone sensor nodes. The network is built and maintained by the backbone nodes such that a backbone node can communicate with the network sink over a multi-hop path using other backbone nodes as routers. Sensing is done by inexpensive sensors that move along ocean current and send the sensing information to the nearest backbone node. Each backbone node is equipped with a localization component while mobile sensors are oblivious of their locations. There are several challenges to build such a network. Since the backbone nodes are expensive, we want to minimize the number of backbone nodes while ensuring the mobile sensors can have a backbone node within an acceptable distance (i.e., maintaining *coverage*) and also a backbone node can directly communicate with its neighboring backbone nodes (i.e., maintaining *connectivity*). It is also important to make sure that interference at backbone nodes and at mobile sensors is limited.

*Assumptions*

- Homogeneous sphere-based communication: We assume a spherical communication model where any two backbone nodes can communicate if distance between them is less than or equal to a deterministic threshold, say $r_{bb}$, and communication between a backbone node and a mobile sensor can occur if distance between them is less than or equal to another deterministic threshold, say $r_{bs}$.

- No boundary effect: The network is assumed to be very large and there is no boundary effect, so that the number of backbone nodes required to cover the network space is inversely proportional to the volume of a Voronoi cell created by those nodes.

- Large number of inexpensive sensors: We assume that the inexpensive sensors are densely deployed such that any point of the network can be sensed by at least one sensor at any time. As a result sensing coverage is not an issue, rather sending the information back to sink is the challenge. The idea is to use backbone nodes to do that job.

- Adjustable backbone node position: It is assumed that a backbone node can be deployed in any position (or, move to that position) as required by the positioning algorithm. One major criticism of this assumption is that GPS does not work underwater and we do not have any robust positioning mechanism for underwater network yet. However, this assumption ensures that our solution provides the lower bound of the number of nodes needed to achieve full coverage and connectivity. Research in underwater localization is gaining momentum [8] and it is possible that robust underwater positioning mechanism will be available in near future.

### 3.1 Network Topology

In this subsection, we investigate the problem of finding a placement strategy that deploys minimum number of backbone nodes in our hierarchical network such that any mobile sensor can directly communicate with at least one backbone node and any backbone node can directly communicate with any other backbone node, possibly over a multi-hop path, through the network created by the backbone nodes. This problem can be analyzed from the point of view of the shape of virtual Voronoi cells corresponding to the placement of backbone nodes in 3D space [2][3]. If each Voronoi cell is identical and the boundary effect is negligible, then total number of backbone nodes required is equal to the ratio of the volume of the 3D space to be covered to the volume of one Voronoi cell. So minimizing the number of nodes can be achieved if the corresponding virtual Voronoi cell has the highest volume among all placement strategies subject to the constraint that the radius of its circumsphere can not exceed $r_{bs}$. Since achieving the highest volume is the goal, the radius of circumsphere must always be equal to $r_{bs}$ and so the volumes of the circumspheres of all Voronoi cells are the same and equal to $4\pi r_{bs}^3/3$. So our problem reduces to the problem of finding the space-filling polyhedron which has the highest ratio of its volume to the volume of its circumsphere. We call this ratio the *volumetric quotient* of the space-filling polyhedron. Since the volume of the circumsphere is the upper bound on the volume of any polyhedron, the value of volumetric quotient is always between 0 and 1. So our problem is essentially finding the space-filling polyhedron that has the highest volumetric quotient. One possible approach is to check all possible space-filling polyhedrons and determine which space-filling polyhedron has the highest volumetric quotient. However, as shown in [3], a rigorous proof that considers all possible space-filling polyhedrons is difficult to find given the similarity between this problem and Kelvin's conjecture, a century old problem that is still open for identical cells. This problem has been discussed in detail in [3] and in this paper, we restate the results without going into the details. Then we apply these results to build our hierarchical network.

### 3.1.1 Analysis

We start with four different models, namely *CB*, *HP*, *RD* and *TO* model, where shape of the virtual Voronoi cell is respectively, cube, hexagonal prism with the height optimized to maximize the volume of the cell, rhombic dodecahedron and truncated octahedron. As shown in [2], volumetric quotients of cube, hexagonal prism, rhombic dodecahedron and truncate octahedron are $2/\sqrt{3}\pi = 0.36755$, $3/2\pi = 0.477$, $3/2\pi = 0.477$ and $24/5\sqrt{5}\pi = 0.68329$, respectively. So *CB*, *HP* and *RD* model respectively require 85.9%, 43.25% and 43.25% more backbone nodes than the *TO* model. Clearly, the number of backbone nodes needed in *TO* model is significantly smaller than the other three models. Now, if we consider the connectivity among backbone nodes, then none of the above model works for all values of $r_{bb}/r_{bs}$. In order to keep any two physically neighboring backbone nodes within the value of $r_{bb}$, *CB*, *HP*, *RD* and *TO* model requires that the value of $r_{bb}/r_{bs}$ is at least $2/\sqrt{3} = 1.1547$, $\sqrt{2} = 1.4142$, $\sqrt{2} = 1.4142$ and $4/\sqrt{5} = 1.7889$, respectively. We adjust the models as follows to maintain connectivity of a backbone node with all neighboring backbone nodes for all values of $r_{bb}/r_{bs}$. In the case of *Adjusted CB*, *RD* and *TO* model, we set the radius of circumsphere to be $R = \min(\sqrt{3}r_{bb}/2, r_{bs})$, $R = \min(r_{bb}/\sqrt{2}, r_{bs})$, $R = \min(r_{bb}\sqrt{5}/4, r_{bs})$, respectively. In the case of *HP*, we set each side of the hexagon to $a = \min(r_{bb}/\sqrt{3}, r_{bs}\sqrt{2}/\sqrt{3})$ and the height of each hexagonal prism to $h = \min(2\sqrt{r_{bs}^2 - a^2}, r_{bb})$. Then the volume of a cell in all four models for all values of $r_{bb}/r_{bs}$ can be easily determined and is shown in Figure 1.

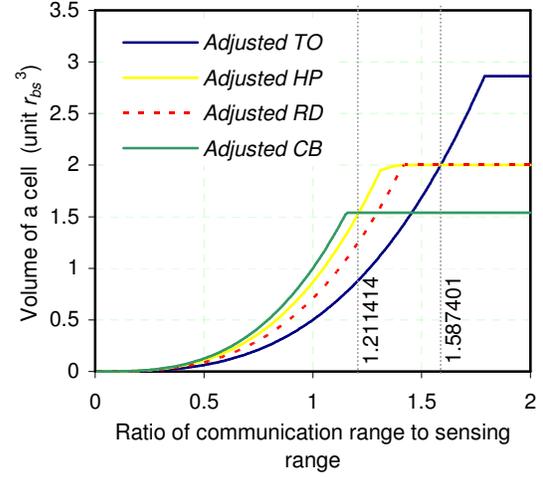

Figure 1: Comparison of different placement models for hierarchical network

From the figure, we see that when $r_{bb}/r_{bs} \geq 1.587401$, we can use the *Adjusted TO* model; when $1.587401 > r_{bb}/r_{bs} \geq 1.211414$, *Adjusted HP* is the best option and when $r_{bb}/r_{bs} < 1.211414$, *Adjusted CB* has the best performance. Backbone node placement in *CB*, *HP*, *TO* and *RD* model can be achieved by taking any arbitrary point (*x, y, z*) as a reference and deploying one backbone node in co-ordinates $\left(x + u\frac{2R}{\sqrt{3}}, y + v\frac{2R}{\sqrt{3}}, z + w\frac{2R}{\sqrt{3}}\right)$, $\left(x + u \times a\sqrt{3}\sin 60^0, y + u \times a\sqrt{3}\cos 60^0 + v \times a\sqrt{3}, z + w \times h\right)$, $\left(x + (2u+w)\frac{R}{\sqrt{2}}, y + (2v+w)\frac{R}{\sqrt{2}}, z + wR\right)$ and $\left(x + (2u+w)\frac{2R}{\sqrt{5}}, y + (2v+w)\frac{2R}{\sqrt{5}}, z + w\frac{2R}{\sqrt{5}}\right)$ where $u \in \mathbb{Z}, v \in \mathbb{Z}, w \in \mathbb{Z}$; $\mathbb{Z}$ is the set of integers (both negative and positive). Clearly, this approach deploys nodes in the entire 3D Euclidean space and creates a network that is infinite along all three axes. In practice, 3D networks will be finite and backbone node placement can be made for a finite network by considering only those co-ordinates that fall within the space that we want to cover. Same discussion applies to other contexts in this paper wherever $\mathbb{Z}$ is used.

When $r_{bb}/r_{bs} \geq 4/\sqrt{5}$, ensuring full coverage (i.e., any mobile sensors has at least one backbone node within $r_{bs}$ distance) with minimum number of backbone nodes automatically ensures all backbone nodes are connected with all 14 of their physically neighboring nodes (i.e., full connectivity) in the original *TO* model. So the overhead for full connectivity is zero. However, when the value of $r_{bb}/r_{bs}$ is small, a significant number of extra nodes has to be deployed to ensure full connectivity even after full coverage is already achieved. If we relax the requirement of full connectivity, then communication among distant nodes in general takes longer route and if nodes are failure prone, there is a chance that some nodes may be totally disconnected. So here we have a trade off between faster communication and the number of nodes needed. Relaxing full connectivity with all first tier neighboring nodes makes sense when the nodes are expensive and robust with very low probability of failure, and the value of $r_{bb}/r_{bs}$ is small. For example, when $2\sqrt{3}/\sqrt{5} = 1.549193 \leq r_{bb}/r_{bs} < 4/\sqrt{5}$, in original *TO* model, a backbone node still has direct connectivity with 8 neighboring backbone nodes (see Figure 2 and Figure 3), which may be sufficient if backbone nodes are robust.

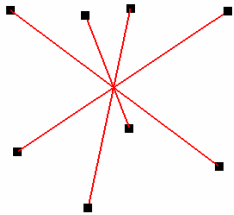

Figure 2: A backbone node has links with 8 neighboring backbone nodes in a *TO* based hierarchical network when $2\sqrt{3}/\sqrt{5} \leq r_{bb}/r_{bs} < 4/\sqrt{5}$. This number increases to 14 for higher values of $r_{bb}/r_{bs}$.

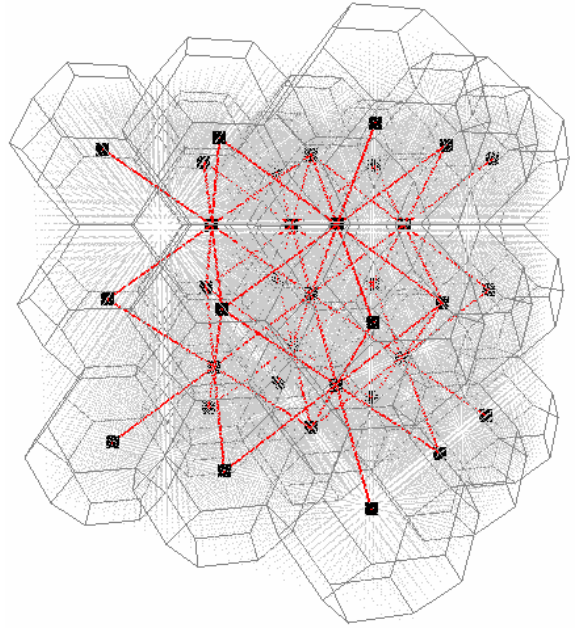

Figure 3: A TO based hierarchical network when $2\sqrt{3}/\sqrt{5} \leq r_{bb}/r_{bs} < 4/\sqrt{5}$. Small grey dots are failure prone mobile sensors. Large black dots are backbone nodes. Backbone network links are shown with red lines. When $r_{bb}/r_{bs} \geq 4/\sqrt{5}$, each inner backbone node has 14 links as opposed to 8 links shown in the figure.

For even smaller values of $r_{bb}/r_{bs}$, we can decrease the number of backbone nodes even more by using the following strip based node placement strategy that can provide full coverage and 1-connectivity[1] [3]. Deploy nodes as strips such that the distance between any two nodes in a strip as $\alpha = \min\{r_{bb}, 4r_{bs}/\sqrt{5}\}$ and keep distance between two parallel strips in a plane as $\beta = 2\sqrt{r_{bs}^2 - (\alpha/4)^2}$. Set distance between two planes of strips as $\beta/2 = \sqrt{r_{bs}^2 - (\alpha/4)^2}$ and deploy strips such that a strip of one plane is placed between two nearest strips of a neighboring plane. Here distance between two neighboring nodes that reside in two different planes is $\gamma = \sqrt{\beta^2/2 + \alpha^2/4}$. This deployment of sensors can be achieved by taking a reference point (*x*, *y*, *z*) and placing a node at each of the coordinates $(x + u\alpha + w\gamma\cos\theta, y + v\beta + w\gamma\cos\theta, z + w\gamma\cos\theta)$, where $\theta = \cos^{-1}(\sqrt{1/3})$ and $u \in \mathbb{Z}, v \in \mathbb{Z}, w \in \mathbb{Z}$.

---

[1] We say that our hierarchical network has *k*-connectivity if every backbone node can communicate with every other backbone nodes of the network along at least *k* different paths.

Unless $\beta \leq r_{bb}$ or, $\gamma \leq r_{bb}$, this strip based approach only ensures connectivity among nodes in the same strip. In order to ensure connectivity between strips, we need to place additional nodes between strips. We can achieve 1-connectivity by placing auxiliary nodes such that any two neighboring nodes in two strips are connected. However, 2-connectivity can only be achieved by placing auxiliary nodes at the two endpoints of the strips along the boundary of the network. Unless $\beta \leq r_{bb}$ or, $\gamma \leq r_{bb}$, there is no way to achieve 3- or higher connectivity without deploying a large number of auxiliary nodes.

So we can deploy backbone nodes and build the backbone network according to above guideline. In the case of mobile sensors, all we need is to make sure that they are densely deployed in the network space such that sensing coverage is always maintained. Any particular mobile sensor does not have to be in any particular place. It may move by ocean current etc. Since we deploy the backbone nodes in such a way that there is at least one backbone node within an acceptable distance of the mobile node, it can always communicate with the backbone node and thus to the network sink.

### 3.2 Communication among nodes

In a terrestrial cellular network, base stations usually communicate among themselves through wired network and communication between a base station and a mobile subscriber is through wireless medium. In our hierarchical network, communication among backbone nodes and communication between a backbone node and a mobile sensor can also different communication media. For example, acoustic communication can be used between two backbone nodes while optical communication may be used between a backbone node and a mobile sensor. Still there will be interference both at backbone nodes and at mobile sensors if all backbone nodes communicate with all mobile sensors over the same channel. So we need to make sure that backbone nodes and mobile sensors communicating over same channel are sufficiently far apart. This requires us to apply frequency reuse concept widely used in terrestrial cellular network with the exception that the network is now three-dimensional.

Spatial frequency reuse in cellular networks has been a key technological breakthrough in solving the scarcity of frequency spectrum in terrestrial mobile radio communication. Since available bandwidth is very limited in underwater acoustic environment, spatial frequency reuse in underwater acoustic networks is a very promising idea [23]. However, following two fundamental differences between a terrestrial cellular radio network and an underwater acoustic network limit the direct application of existing results and so new research need to be done for spatial frequency reuse in an underwater acoustic network:

1. Well known hexagon based solution for terrestrial 2D cellular networks is not applicable in a 3D underwater acoustic network.

2. In a terrestrial radio network, the signal power attenuates with distance as $P(d) \sim 1/d^n$ where *n* is the path loss exponent whose value is usually between 2 and 4 and *d* is the distance traveled by the signal. On the other hand, in an underwater acoustic environment path loss depends on frequency in addition to the distance traveled. More formally, the path loss experienced by a acoustic signal of frequency *f* traveling over a distance *d* is given by $A(d,f) = A_0 d^n a^d(f)$ where $A_0$ is a normalizing constant, *n* is the spreading factor and *a(f)* is the absorption coefficient that depends on the frequency [23].

The difference in path loss has been investigated in [24], but that work still assumes that the nodes are deployed on a 2D plane and uses the hexagon based solution of terrestrial cellular radio network for the analysis. In the next subsection, we address the first difference by investigating the frequency reuse for a 3D network but do not consider the second difference. Then in the next subsection, we consider addressing both differences jointly.

*Frequency Reuse and Clusters in 3D Radio Network*

Here we show how frequency reuse can be done for different kind of three-dimensional cells where signal power attenuates with distance as $P(d) \sim 1/d^n$.

- *Rhombic Dodecahedron*: Rhombic dodecahedron tessellation of the 3D space can be formed if the center of each cell is located in the integer coordinates of the following coordinate system consisted of $u$, $v$ and $w$ axes. Unit distance along each axis is $\sqrt{2}R$ where $R$ is the radius of a cell. The positive portions of any two axes form a $60^0$ angle. The distance between any two points $(u_1, v_1, w_1)$ and $(u_2, v_2, w_2)$ is $D = \sqrt{2}R\sqrt{i^2 + j^2 + k^2 + ij + jk + ki}$ where $i = u_2 - u_1$, $j = v_2 - v_1$ and $k = w_2 - w_1$. In order to determine the cluster size, the co-channel cells have to be placed at equidistant points from a reference co-channel cell. If we impose the restriction that the reuse distance must be isotropic, there are 12 rhombic dodecahedra equidistant from the reference rhombic dodecahedron. Assuming that a cluster has a rhombic dodecahedral shape, we want to determine the number of rhombic dodecahedral cells per cluster. Let $v$ and $V$ be the volume of the cell and of the cluster, respectively. The volume $v$ is then $v = 2R^3$. Let us choose the centers of two co-channel cells two be the centers of the corresponding rhombic dodecahedral clusters. Then the volume $V$ is

$$V = 2\left[\frac{1}{\sqrt{2}}\sqrt{2}R\sqrt{i^2 + j^2 + k^2 + ij + jk + ki}\right]^3$$

$$= 2R^3\left(i^2 + j^2 + k^2 + ij + jk + ki\right)^{\frac{3}{2}}.$$

So the number of cells per cluster is $N = \frac{V}{v} = \left(i^2 + j^2 + k^2 + ij + jk + ki\right)^{\frac{3}{2}}$.

Since $i$, $j$ and $k$ are integers and a logical constraint is that the number of cells per cluster has to be an integer, a cluster can accommodate a certain number of cells, such as, 1, 8, 27 and so on. The co-channel reuse ratio is $D/R = \sqrt{2}N^{\frac{1}{3}}$. Since the number of co-channels is 12, the signal to interference ratio (SIR) is $S_r = \frac{1}{12}\left(\frac{D}{R}\right)^4 = \frac{1}{3}N^{\frac{4}{3}}$.

- *Cube*: Cube tessellation of the 3D space can be formed if the center of each cell is located in the integer coordinates of the Cartesian coordinate system with unit distance along each axis is $2R/\sqrt{3}$. The distance between two points $(u_1, v_1, w_1)$ and $(u_2, v_2, w_2)$ is given by $D = \frac{2}{\sqrt{3}}R\sqrt{i^2 + j^2 + k^2}$ where $i = u_2 - u_1$, $j = v_2 - v_1$ and $k = w_2 - w_1$. In order to determine the cluster, the co-channel cells have to be placed at equidistant points from a reference co-channel cell. Assuming that a cluster has a cubical shape, the number of cube cells per cluster is $N = n^3, \forall n \in \mathbb{N}$. The co-channel reuse ratio is $D/R = 2N^{\frac{1}{3}}/\sqrt{3}$

- *Truncated Octahedron:* Truncated octahedron tessellation of the 3D space can be formed if the center of each cell is located in the integer coordinates of the following coordinate system consisted of three axes $u$, $v$ and $w$. Unit distance in $u$ and $v$ axis is $4R/\sqrt{5}$ and unit distance along $w$ axis is $2\sqrt{3}R/\sqrt{5}$. Angles between the axes are $\angle uv = 90^0$ and $\angle uv = \angle vw = \cos^{-1}\left(\sqrt{1/3}\right) = 54.73^0$. Axis $w$ creates an angle $\sin^{-1}\left(\sqrt{1/3}\right) = 35.264^0$ with the $uv$ plane. In this coordinate system, distance between two points $(u_1, v_1, w_1)$ and $(u_2, v_2, w_2)$ is given by $D = \frac{4}{\sqrt{5}}R\sqrt{i^2 + j^2 + ik + jk + \frac{3}{4}k^2}$ where $i = u_2 - u_1$, $j = v_2 - v_1$ and $k = w_2 - w_1$. The volume of a cell is $v = 32R^3/5\sqrt{5}$. Assuming that a cluster has a truncated octahedral shape and by choosing the centers of two co-channel cells as the centers of the corresponding rhombic dodecahedral clusters, we have the volume of a cluster to be

$$V = \frac{32}{5\sqrt{5}} \left[ \frac{1}{\frac{4}{\sqrt{5}}} \frac{4}{\sqrt{5}} R \sqrt{i^2 + j^2 + ik + jk + \frac{3}{4}k^2} \right]^3$$

$$= \frac{32}{5\sqrt{5}} R^3 \left( i^2 + j^2 + ik + jk + \frac{3}{4}k^2 \right)^{\frac{3}{2}}$$

So the number of cells per cluster is $N = \frac{V}{v} = \left( i^2 + j^2 + ik + jk + \frac{3}{4}k^2 \right)^{\frac{3}{2}}$. Since $i, j$ and $k$ are integers and a logical constraint is that the number of cells per cluster has to be an integer, a cluster can accommodate a certain number of cells, such as, 1, 8, 27 and so on. The co-channel reuse ratio is $D/R = \frac{4}{\sqrt{5}} N^{\frac{1}{3}}$.

- *Hexagonal Prism*: Hexagonal prism tessellation of the 3D space can be formed if the center of each cell is located in the integer coordinates of the following coordinate system consisted of $u$, $v$ and $w$ axes. Unit distances along both $u$ and $v$ axes are $\sqrt{2}R$ and unit distance along $w$ axis is $2R/\sqrt{3}$ where $R$ is the radius of a cell. The positive portions of $u$ and $v$ axes form a $60^0$ angle and $w$ axis is orthogonal to $uv$ plane. The distance between two points $(u_1, v_1, w_1)$ and $(u_2, v_2, w_2)$ is given by

$$D = \sqrt{2}R \sqrt{i^2 + j^2 + \frac{2}{3}k^2 + ij} \quad \text{where}$$

$i = u_2 - u_1$, $j = v_2 - v_1$ and $k = w_2 - w_1$.

*Frequency Reuse in 3D Acoustic Network*

Frequency reuse for acoustic network in 2D context has been investigated in [23] and [24]. In this subsection, we update that for 3D networks using the results we obtain in the pervious subsection.

Since [24] is written in the context of 2D networks, it uses the hexagon based model and for their analysis assumes the value of *N*=7 and $SIR = P(R)/6P(D)$. However, for 3D, we may not have all co-channel cells at equal distance from a cell. So we need to use the following more general formula: if the number of co-channel cells is *N* then signal to interference ratio can be defined as follows $SIR = P(R) \Big/ \sum_{i=1}^{N} P(D_i)$, where $D_i$ is the distance traveled by the interfering signal from *i*-th co-channel cell and *R* is the radius of the cell. Using the fact that signal power attenuates with distance as $P(d) \sim 1/d^k$ and in 2D, the reuse factor is $Q = \frac{D}{R} = \sqrt{3N}$, so [24] uses $SIR = Q^k/6$. Path loss of an acoustic signal of frequency *f* traveling over a distance *d* is given by

$$A(d, f) = A_0 d^k a^d(f)$$

where $A_0$ is a normalizing constant, *k* is the spreading factor (the values 1 and 2 corresponds to cylindrical and spherical spreading, respectively), and *a(f)* is the absorption coefficient.

The signal power at a distance *d* from the transmitter is then evaluated as

$$P(d) = \int_{f_n}^{f_n + B_0} S(f) A^{-1}(d, f) df$$

where $S(f) = P_T/B_0$ is the power spectral density of the transmitted signal, which is assumed to be flat and the integration is carried over the frequency occupied by the signal, starting at some $f_n$ and extending over a bandwidth $B_0$.

In this paper, we do our analysis assuming the shape of the cell is rhombic dodecahedron. Analysis of other shapes of the cell should also be similar.

In the case of rhombic dodecahedron shaped 3D cell,

$$SIR = \frac{P(R)}{12P(D)} = \frac{P(R)}{12P\left(\sqrt{2}N^{\frac{1}{3}}R\right)}$$

$$= \frac{\int \frac{S(f)}{A_0 R^k a^R(f)} df}{12\int \frac{S(f)}{A_0 \left(\sqrt{2}N^{\frac{1}{3}}R\right)^k a^{\sqrt{2}N^{\frac{1}{3}}R}(f)} df}$$

$$i.e., SIR = \frac{1}{12}\left(\sqrt{2}N^{\frac{1}{3}}\right)^k \frac{\int_{f_{min}}^{f_{min}+B_0} a^{-R}(f) df}{\int_{f_{min}}^{f_{min}+B_0} a^{-\sqrt{2}N^{\frac{1}{3}}R}(f) df}$$

So SIR depends on both cell radius $R$ and reuse number $N$ for given a frequency range [23].

Figure 4 shows SIR for different $N$ and $R$ when shape of the 3D cell is Rhombic Dodecahedron assuming $k=1.5$, $f_{min} = 10$ kHz and $B_0 = 7$ kHz.

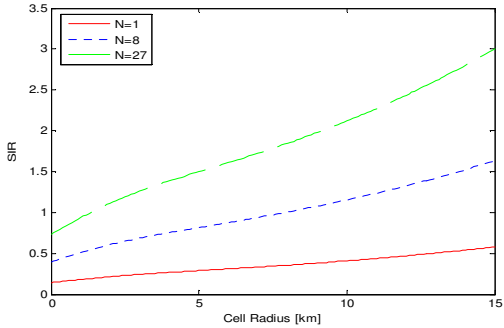

Figure 4: Acoustic SIR for different values of $N$ and $R$ when 3D cell is Rhombic Dodecahedron, $f_{min}=10$ kHz, $B_0=7$ kHz and $k=1.5$.

SIR is strongly affected by the value of $f_{min}$ (Figure 5).

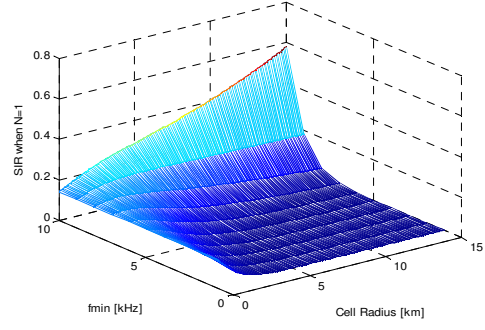
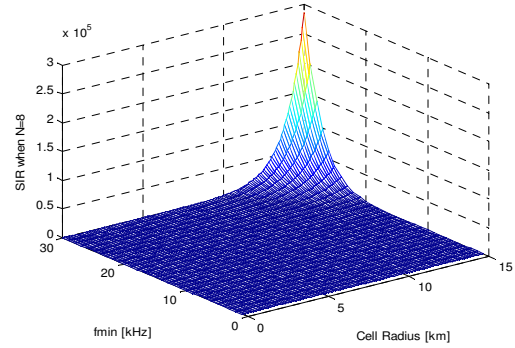

Figure 5: Influence of $f_{min}$ on the value of SIR

Suppose that we have the constraint that per user band width be $W \geq W_0$. Now, if the user density is $\rho$, then the number of users per cell is $2R^3\rho$. If the reuse number is $N$, then bandwidth allocated to each cell is $B/N$, then the constraint is $W = \frac{B/N}{2R^3\rho} \geq W_0$ which implies that the cell radius can not be larger than $\frac{1}{\sqrt[3]{2\rho}}\sqrt[3]{\frac{B}{NW_0}}$. If we impose the restriction that the number of users per cell has to be at least 1, then we have $2R^3\rho \geq 1$, i.e., $R \geq \frac{1}{\sqrt[3]{2\rho}}$. So the cell radius must satisfy the following condition $\frac{1}{\sqrt[3]{2\rho}} \leq R \leq \frac{1}{\sqrt[3]{2\rho}}\sqrt[3]{\frac{B}{NW_0}}$.

Another restriction on $R$ is set by constraint of minimum SIR (say, SIR$_0$) which is

$$SIR_0 \leq \frac{1}{12}\left(\sqrt{2}N^{\frac{1}{3}}\right)^k \frac{\int_{f_{min}}^{f_{min}+B_0} a^{-R}(f) df}{\int_{f_{min}}^{f_{min}+B_0} a^{-\sqrt{2}N^{\frac{1}{3}}R}(f) df}$$

Once the reuse number *N* is fixed, the cell radius *R* can be chosen that maximizes the number of users.

### 3.3 Energy Consumption

Here we compare energy efficiency for our four original models, namely *CB, HP, RD, TO* model. We assume that all mobile sensors are always on and they send packet to nearest backbone node with same signal strength that achieves the distant requirement of $r_{bs}$. As a result, difference in energy efficiency in different models comes from different energy requirements of backbone nodes. We assume a backbone node uses different signal strength in different model such that transmission range is equal to the distance between two neighboring nodes in that particular model. It is also assumed that power consumption is primarily due to communication and difference in energy requirement in different model depends on the transmission range used by a backbone node.

At first we compare relative energy requirements to send a packet to the sink over multi-hop path in each model. If distance between two neighboring backbone nodes is $r_{bb}$, then for each packet generated at distance *D* from the sink, total number of intermediate hops plus the source backbone nodes (i.e., number of transmissions) is $\lceil D/r_{bb} \rceil$. For simplicity of calculation we use $D/r_{bb}$ instead which is a reasonable approximation for large *D* and small *r*. Now, for two models with transmission range for backbone nodes as $r_{bb'}$ and $r_{bb''}$, per packet power consumption ratio in each hop is $\frac{P^{h'}}{P^{h''}} = \frac{r_{bb'}^2}{r_{bb''}^2}$. So to send each packet all the way to the sink, the power consumption ratio in two models is $\frac{P'}{P''} = \frac{r_{bb'}}{r_{bb''}}$. Following table shows power consumption ratio of each of the four models with respect to *TO* model.

Table 1: Power consumption ratio per packet for each model with respect to *TO* model

| Model | Power consumption ratio per packet |
|---|---|
| CB | 0.64548 |
| HP | 0.79054 |
| RD | 0.79054 |
| TO | 1.00000 |

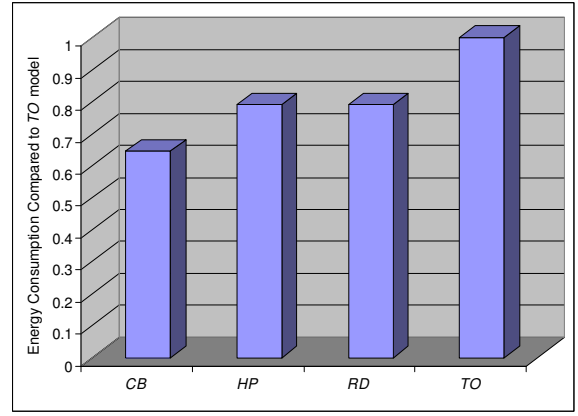

Figure 6: Per packet energy consumption comparison among various models

Now if we assume that total number of packet generated by each model is same, then clearly *CB* model has the smallest power consumption which is obvious given the well known fact that the lower the transmission range, the lower the power consumption. However, this answer is misleading, given that we are not considering cost associated with increase number of nodes used by CB model (85.9% more nodes than *TO* model).

An alternative model can assume that each source node can aggregate information and send one packet irrespective of the number of mobile sensors it covers, i.e., the number of packets generated by each model is proportional to the number of cells in that model. Then ratio of power consumption by the entire network in each model is essentially power consumption ratio per packet times the ratio of the number of cells in each model. Power consumption by entire network in each model with respect to *TO* model is shown in the following table.

Table 2: Power consumption ratio of entire network in each model with respect to *TO* model

| Model | Power consumption ratio of entire network |
|---|---|
| CB | 0.64548×1.859 = 1.1999 |
| HP | 0.79054×1.4325=1.1325 |
| RD | 0.79054×1.4325=1.1325 |
| TO | 1.0000 |

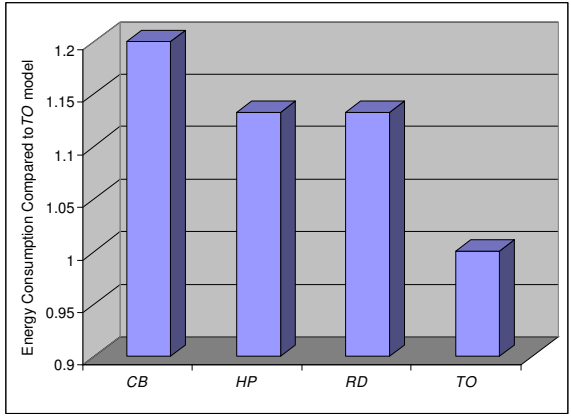

Figure 7: Total energy consumption comparison among various models

Power consumption per backbone node is highest in *TO* model which is reasonable, because *TO* model deploys far fewer backbone nodes than any other model, and as a result it must place backbone nodes further apart which leads to higher transmission range. However, when we take into account the number of backbone nodes deployed in each model by comparing power consumption by the entire network, *TO* is model is the most energy efficient among the four models.

## 4. NONHIERARCHICAL NETWORK ARCHITECTURE

In this section, we consider the scenario where deploying and maintaining a carefully planned backbone network is not feasible. This may happen if only one type of sensor nodes is available and the sensor nodes cannot be deployed in pre-determined positions and/or they cannot maintain predetermined positions due to ocean current, gravity, fish and other marine animals etc. So the topology control algorithm has to assume that the sensor nodes are randomly deployed. However, due to this random deployment, full coverage and connectivity can be ensured if a lot of redundant nodes are densely deployed. However, keeping redundant nodes active increases the consumption of valuable energy and also may increase congestion by sending redundant messages. So it is important to find a dynamic mechanism that decreases the redundant active nodes by selecting a subset of the nodes to act as active nodes in a dynamic and distributed fashion in real time. One simple way to do that is to partition the network space into cells and keep one node active in each cell. In order to make the selection process distributed, we also impose the restriction that cells are identical. Clearly, the smaller the number of active nodes at a time, the higher the energy saving. However, maintaining full connectivity requires that the maximum distance between the active nodes of any two first-tier neighboring cells cannot exceed the transmission radius (a.k.a. communication range). Since the active node can be located anywhere inside a cell, the maximum distance between any two points of two first-tier neighboring cells must be less than or equal to the transmission radius. One major work in this context is geographic adaptive fidelity (GAF) [29]. Although GAF is proposed to maintain fidelity in routing of a 2D wireless ad hoc network, the concept can easily be extended to a 2D wireless sensor network to maintain fidelity in coverage and connectivity. GAF divides a 2D network into squared virtual cells (a.k.a. grids) and keeps one node active in each cell. It can be shown that GAF performs better when the shape of a virtual cell is a hexagon instead of a square. The energy savings in GAF depends heavily on the choice of the partitioning scheme, because the number of active nodes at a time is equal to the number of total virtual cells. Clearly, hexagonal partitioning scheme of 2D networks is not applicable in 3D networks. In this section, we investigate and provide a solution for this partitioning problem in 3D. In particular, we have the following assumptions and goals.

*Assumptions*

- The sensors are uniformly and densely distributed over a 3D space.

- All sensor nodes are identical. For example, they have identical fixed transmission range $r_t$ and identical energy source (battery). Transmission is omni-directional and the transmission region of each node can be represented by a sphere of radius $r_t$, having the node at its center.

- The transmission range $r_t$ is much smaller than the length, the width, or the height of the 3D space to be covered, so that the boundary effect is negligible and hence can be ignored.

- There is a localization component in each sensor node that allows it to determine its location in the 3D space.

*Goals*

- Given any fixed transmission range $r_t$, find the best partitioning scheme to divide a 3D space into identical virtual cells such that total number of virtual cells are minimum over all possible virtual schemes and the maximum distance between any two points of two neighboring virtual cells does not exceed the transmission range $r_t$. We keep the transmission range fixed for all models, so that the energy consumption due to transmission remains same and we have a fair comparison.
- Find the minimum sensing range in terms of transmission radius such that any two points in a virtual cell does not exceed the sensing range.
- Find an algorithm so that such partitioning (i.e., each sensor node knows in which cell they belong) can be made in a fast, efficient and distributed manner.
- We also want to know how efficient the scheme is as compared to a scheme where an oracle determines which nodes to keep active. We want it for both 2D GAF and our scheme in 3D.

It should be noted that any criticism of GAF in 2D also applies to our scheme in 3D. For example, even the best possible partitioning scheme may require more than optimal number of active nodes to achieve full coverage and connectivity. However, our approach is decentralized and any scheme that always achieves full coverage and connectivity with minimum number of active nodes needs a centralized scheduling approach which is not always feasible for a large network. Our scheme treats all nodes in a virtual cell as equivalent for coverage and connectivity point of view, and it works well only in a network where nodes are densely and uniformly deployed. In a network where no node is physically located in a cell, it is no longer true that selecting any node to be active in each cell does not make any difference [6].

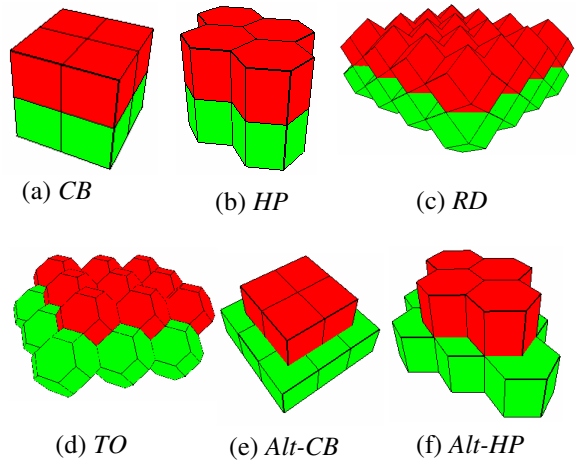

(a) *CB*  (b) *HP*  (c) *RD*

(d) *TO*  (e) *Alt-CB*  (f) *Alt-HP*

Figure 8: 3D Partitioning Schemes

Like in hierarchical network, our focus here is on the four most common polyhedrons that tessellate a 3D space: cube, hexagonal prism, rhombic dodecahedron and truncated octahedron. However, unlike hierarchical network, here the arrangement of cells is also important as the distance between any two points of two neighboring cells must be within the transmission radius. For truncated octahedron and rhombic dodecahedron only one arrangement of cells is possible, the regular 3D space tessellation. On the other hand, for cube and hexagonal prism, an alternate arrangement of cells is possible that asymptotically requires fewer nodes than regular 3D space tessellation. We call these alternate arrangements of cube and hexagonal prism as *Alt-CB* and *Alt-HP* (See Figure 8).

The regular 3D space tessellation of cube, hexagonal prism, rhombic dodecahedron and truncated octahedron shaped cells are referred to as *CB*, *HP*, *RD* and *TO* model, respectively.

## 4.1 Analysis

In this subsection, we briefly analyze all six models. Given a fixed transmission radius $r_t$, the maximum radius of a cell in *CB*, *Alt-CB*, *HP*, *Alt-HP, RD* and *TO* models is calculated below.

*CB model*: A cell has 26 first tier neighboring cells: 6 *Type* $1_{CB}$ neighboring cells each share whole one side of a cube, 12 *Type* $2_{CB}$ neighboring cells each share a common line and 8 *Type* $3_{CB}$ neighboring cells each share just a common point with the cell (See Figure 9).

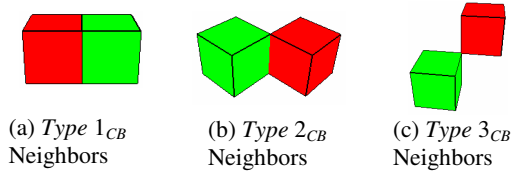

(a) *Type 1$_{CB}$* Neighbors  (b) *Type 2$_{CB}$* Neighbors  (c) *Type 3$_{CB}$* Neighbors

Figure 9: Different types of neighbors in *CB* model

Suppose that the radius of a cube is *R*. Then the largest distance between any point in the cell and any point in a *Type 1$_{CB}$* neighboring cells is $R2\sqrt{2}$ =2.828427$R$; for *Type 2$_{CB}$* and *Type 3$_{CB}$* neighbors, it is $R2\sqrt{3}$ =3.4641$R$ and 4$R$, respectively. So the active node of a cell can communicate with active nodes of all first-tier neighboring cells if the maximum radius of a cell in *CB* model is $r = \frac{r_t}{\max(2\sqrt{2}, 2\sqrt{3}, 4)} = \frac{r_t}{4} = 0.25 r_t$.

*Alt-CB model*: A cell has 16 first tier neighboring cells: 4 *Type 1$_{Alt-CB}$* neighboring cells each share whole one side of a cube, 4 *Type 2$_{Alt-CB}$* neighboring cells each share a common line, and 8 *Type 3$_{Alt-CB}$* neighboring cells each share one quarter of one side of the cell (See Figure 10).

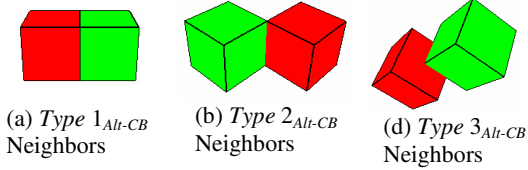

(a) *Type 1$_{Alt-CB}$* Neighbors  (b) *Type 2$_{Alt-CB}$* Neighbors  (d) *Type 3$_{Alt-CB}$* Neighbors

Figure 10: Different types of neighbors in *Alt-CB* model

The largest distance for *Type 1$_{Alt-CB}$* , *Type 2$_{Alt-CB}$* and *Type 3$_{Alt-CB}$* cells is $R2\sqrt{2}$, $R2\sqrt{3}$, and $R\sqrt{34/3}$, respectively. So the maximum radius of an *Alt-CB* cell is $r = \frac{r_t}{\max(2\sqrt{2}, 2\sqrt{3}, \sqrt{34/3})} = \frac{r_t}{2\sqrt{3}} = 0.288675 r_t$

*HP model:* A cell has 20 first tier neighboring cells: 6 *Type 1$_{HP}$* neighboring cells each share a common square plane and 2 *Type 2$_{HP}$* neighboring cells each share a common hexagonal plane and 12 *Type 3$_{HP}$* neighboring cells each share a common line with the cell (See Figure 11).

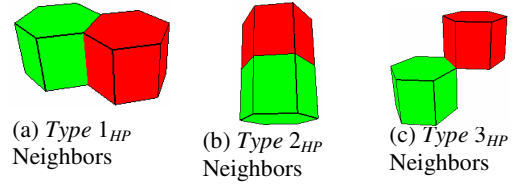

(a) *Type 1$_{HP}$* Neighbors  (b) *Type 2$_{HP}$* Neighbors  (c) *Type 3$_{HP}$* Neighbors

Figure 11: Different types of neighbors in *HP* model

Suppose that each side of a hexagonal face of a *HP* cell is of length *a*, and its height is *h*. In a *HP* cell with optimal height, $h = a\sqrt{2}$. So the radius of *HP* cell is $R = \sqrt{a^2 + \frac{a^2}{2}} = a\sqrt{\frac{3}{2}}$. So maximum distance from any point of the cell to any point of a *Type 1$_{HP}$, Type 2$_{HP}$* and *Type 3$_{HP}$* neighbor is $\sqrt{(a\sqrt{13})^2 + h^2} = R\sqrt{10}$, $\sqrt{(2a)^2 + (2h)^2} = R\sqrt{8}$ and $\sqrt{(a\sqrt{13})^2 + (2h)^2} = R\sqrt{14}$, respectively. So the active node of a cell can communicate with active nodes of all neighboring cells if the maximum radius of a cell in *HP* model is $r = \frac{r_t}{\max(\sqrt{10}, \sqrt{8}, \sqrt{14})} = \frac{r_t}{\sqrt{14}} = 0.26726 r_t$

*Alt-HP model:* A cell has 12 first-tier neighboring cells: 6 *Type 1$_{Alt-HP}$* neighboring cells each share a square plane and 6 *Type 2$_{Alt-HP}$* neighboring cells each share one third of a hexagonal plane with the cell (See Figure 12).

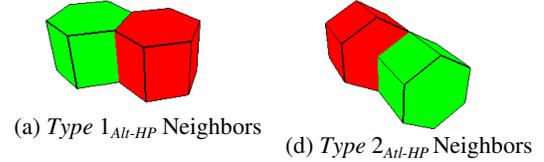

(a) *Type 1$_{Alt-HP}$* Neighbors  (d) *Type 2$_{Atl-HP}$* Neighbors

Figure 12: Different types of neighbors in Alt-HP model

Maximum distance for *Type 1$_{Alt-HP}$* and *Type 2$_{Alt-HP}$* neighbors is $\sqrt{(a\sqrt{13})^2 + h^2} = R\sqrt{10}$ and $\sqrt{(3a)^2 + (2h)^2} = R\sqrt{34/3}$, respectively. So maximum radius of is a cell in *Alt-HP* model is $r = \frac{r_t}{\max(\sqrt{10}, \sqrt{34/3})} = \frac{r_t}{\sqrt{34/3}} = 0.297 r_t$.

*RD model*: A cell has 18 first tier neighboring cells: 6 *Type 1$_{RD}$* neighboring cells each share just a point and 12 *Type $2_{RD}$* neighboring cells each share a plane with the cell (See Figure 13).

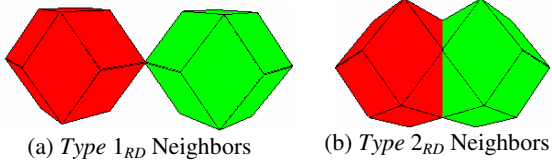

(a) *Type $1_{RD}$* Neighbors     (b) *Type $2_{RD}$* Neighbors

Figure 13: Different types of neighbors in *RD* model

Maximum distance for *Type $1_{RD}$* and *Type $2_{RD}$* neighbor is $4R$ and $R\sqrt{10}$, respectively. So the maximum radius of a cell in *RD* model is $r = \frac{r_t}{\max(4,\sqrt{10})} = \frac{r_t}{4} = 0.25r_t$.

*TO model:* A cell has 14 first tier neighboring cells: 6 *Type $1_{TO}$* neighboring cells each share a common square plane and 8 *Type $2_{TO}$* neighboring cells each share a common hexagonal plane with the cell (See Figure 14).

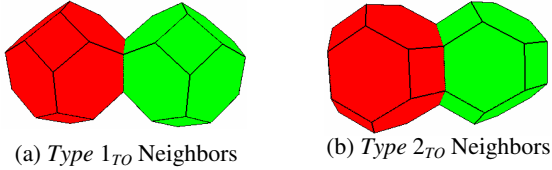

(a) *Type $1_{TO}$* Neighbors     (b) *Type $2_{TO}$* Neighbors

Figure 14: Different types of neighbors in *TO* model

Maximum distance for *Type $1_{TO}$* and *Type $2_{TO}$* neighbor is $\frac{2R}{\sqrt{5}}\sqrt{17}$ and $\frac{2R}{\sqrt{5}}\sqrt{14}$, respectively. So the active node of a cell can communicate with active nodes of all neighboring cells if the maximum radius of a cell in *TO* model is

$$r = \frac{r_t}{\max\left(\frac{2\sqrt{17}}{\sqrt{5}}, \frac{2\sqrt{14}}{\sqrt{5}}\right)} = \frac{r_t\sqrt{5}}{2\sqrt{17}} = 0.271163r_t$$

### 4.1.1 Minimum Sensing Range

Since an active node can be located anywhere inside a cell and still it must be able to sense any point inside the cell, the sensing range must be at least equal to the maximum distance between any two points of a cell. This maximum distance is essentially the diameter of a cell and equal to twice of the corresponding radius. So minimum sensing range of a cell in *CB*, *Alt-CB*, *HP*, *Alt-HP*, *RD* and *TO* model is $2r_t/4 = 0.5r_t$, $2r_t/2\sqrt{3} = 0.577r_t$, $2r_t/\sqrt{14} = 0.535r_t$, $2r_t/\sqrt{34/3} = 0.594r_t$, $2r_t/4 = 0.5r_t$ and $2\sqrt{5}r_t/2\sqrt{17} = 0.542r_t$, respectively (See Figure 15).

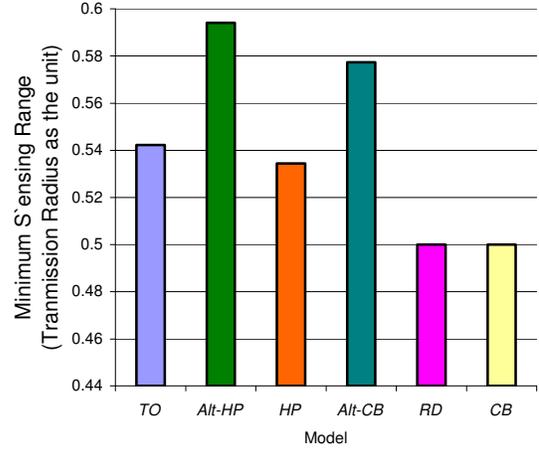

Figure 15: Minimum sensing range in various models

### 4.1.2 Distributed partitioning scheme

The nonhierarchical network architecture can easily be done in a distributed fashion if all nodes know in their cell id. Since the technique is similar for all models, here we provide calculation only for the *TO* model. Suppose that the information sink (*IS*), where all data are gathered, resides in the center of a virtual cell and its coordinate $(x,y,z)$ is known. Then for *TO* model the center of a virtual cell can be expressed by the general equation

$$f(u,v,w) = \left(x + (2u+w)\frac{r_t}{\sqrt{17}}, y + (2v+w)\frac{r_t}{\sqrt{17}}, z + w\frac{r_t}{\sqrt{17}}\right)$$

and three integers $(u,v,w)$ can be used as unique cell id with the cell containing *IS* has the cell id $(0,0,0)$. As an example, cell id $(-1, -1, 2)$ has its center in $(x, y, z + 2r_t/\sqrt{17})$.

Now a sensor node can determine its own coordinate $(x_s, y_s, z_s)$ using its localization component, *IS* can broadcast its coordinate $(x, y, z)$ to all nodes and the transmission radius $r_t$ can be embedded in the sensor before deployment. Now to determine its cell id $(u_s, v_s, w_s)$, a brute force method is to check all possible values of $(u_s, v_s, w_s)$ and choose the cell whose center has minimum Euclidean distance from the node, i.e.,

$$(u_s, v_s, w_s) = \arg\min_{\substack{u \in \mathbb{Z}, \\ v \in \mathbb{Z}, \\ w \in \mathbb{Z}}} \left( x_s - x - (2u+w)\frac{r_t}{\sqrt{17}} \right)^2$$
$$+ \left( y_s - y - (2v+w)\frac{r_t}{\sqrt{17}} \right)^2$$
$$+ \left( z_s - z - w\frac{r_t}{\sqrt{17}} \right)^2$$

where $\mathbb{Z}$ is set of all integers. However, we do not need to do the exhaustive search. Since the value of a square term is never negative, we can set the value of the square terms to zero to get the values of $u_s$, $v_s$ and $w_s$. Since these values must be integer, we can get two possible integral values for each variable by taking ceiling (denoted by subscript $h$) and floor (subscript $l$):

$$u_l = \left\lfloor (x_s - x - z_s + z)\sqrt{17}/2r_t \right\rfloor,$$
$$u_h = \left\lceil (x_s - x - z_s + z)\sqrt{17}/2r_t \right\rceil,$$
$$v_l = \left\lfloor r(y_s - y - z_s + z)\sqrt{17}/2r_t \right\rfloor,$$
$$v_h = \left\lceil (y_s - y - z_s + z)\sqrt{17}/2r_t \right\rceil,$$
$$w_l = \left\lfloor (z_s - z)\sqrt{17}/r_t \right\rfloor, \; w_h = \left\lceil (z_s - z)\sqrt{17}/r_t \right\rceil.$$

Thus we have eight possible values of $(u_s, v_s, w_s)$. Each node has to calculate its distance from each of the eight centers and choose the minimum one as its cell id, i.e.,

$$(u_s, v_s, w_s) = \arg\min_{\substack{u \in \{u_l, u_h\}, \\ v \in \{v_l, v_h\}, \\ w \in \{w_l, w_h\}}} \sqrt{\left( x_s - x - (2u+w)\frac{r_t}{\sqrt{17}} \right)^2 + \left( y_s - y - (2v+w)\frac{r_t}{\sqrt{17}} \right)^2 + \left( z_s - z - w\frac{r_t}{\sqrt{17}} \right)^2}$$

As cell id is a straightforward function of the location of a sensor, if a sensor knows the location of another sensor, it can readily calculate the cell id of that sensor. We use simulation to validate that each sensor node can determine its cell id correctly according to above technique. In a very large number of trials, we found that in every case our equations (ceiling and floor approach) can predict the cell id correctly. However, further effort to simplify the prediction process does not work. For example, Instead of calculating the distance from each of the eight centers, if we simply take the nearest integer value for $u_s$, $v_s$, $w_s$, then this approximation leads to incorrect prediction of cell id in almost one quarter of the cases (See Figure 16).

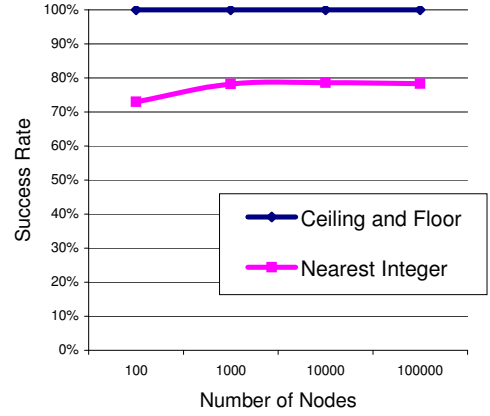

Figure 16: Cell ID prediction accuracy

However, since there are just eight possible combinations, calculation involve in our technique to find the cell id involves just a small constant number of local arithmetic operations.

Once sensors have their cell id, then sensors with same cell id can use any standard leader selection algorithms [18] to choose a leader among them which can act as the active node of that cell. All nodes that have same cell id are within the communication range of each other and the mechanism of keeping one node active among all the sensors with same cell id is essentially same for both 2D and 3D networks. Since the main focus of this paper is problems that are unique to 3D networks, we choose not to explore the issues that have already been studied in the context of 2D networks.

### 4.1.3 Number of Active Nodes and Network Lifetime

Ignoring boundary effect, the number of cells in a network is inversely proportional to the volume of the network. Since at a time the number of active nodes in a cell is one, total number of active nodes in a network is equal to the number of cells in the network. The volume of a cube, hexagonal prism, rhombic dodecahedron and truncated octahedron of radius $R$ is $8R^3/3\sqrt{3}$, $2R^3$, $2R^3$ and $32R^3/5\sqrt{5}$, respectively. Using the maximum radius calculated before, we have the volume of a cell in *CB*, *Alt-CB*, *HP*, *Alt-HP*, *RD* and *TO* models are

$8\left(\dfrac{r_t}{4}\right)^3 / 3\sqrt{3} = \dfrac{r_t^3}{24\sqrt{3}}$, $\quad 8\left(\dfrac{r_t}{2\sqrt{3}}\right)^3 / 3\sqrt{3} = \dfrac{r_t^3}{27}$,

$2\left(\dfrac{r_t}{\sqrt{14}}\right)^3 = \dfrac{r_t^3}{7\sqrt{14}}$, $\quad 2\left(\dfrac{r_t}{\sqrt{34/3}}\right)^3 = \dfrac{3\sqrt{3}r_t^3}{17\sqrt{34}}$,

$2\left(\dfrac{r_t}{4}\right)^3 = \dfrac{r_t^3}{32}$ and $\quad 32\left(\dfrac{r_t\sqrt{5}}{2\sqrt{17}}\right)^3 / 5\sqrt{5} = \dfrac{4}{17\sqrt{17}} r_t^3$,

respectively. So the active nodes required by *CB, Alt-CB, HP, Alt-HP* and *RD* model is, respectively, $96\sqrt{3}/17\sqrt{17}$, $108/17\sqrt{17}$, $28\sqrt{14}/17\sqrt{17}$, $4\sqrt{2}/3\sqrt{3}$ and $128/17\sqrt{17}$ times of that of *TO* model (See Figure 17).

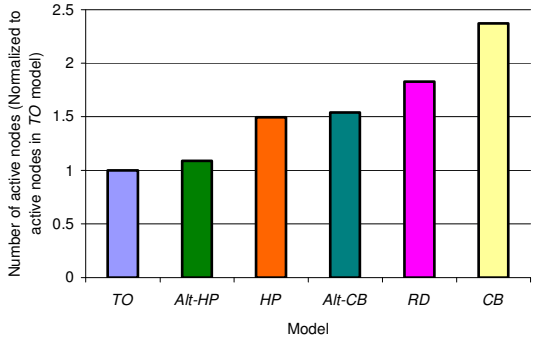

Figure 17: Number of active nodes in various models

Now, we use a simplified model to calculate the network lifetime for different partitioning schemes. Since transmission radius is same in all cases, it can be assumed that a node consumes same amount of power for transmission for all different shapes. If we ignore the power consumption discrepancy due to difference in the number of packets relayed by a node, then the lifetime of an individual node is roughly same in all cases. So lifetime of a cell is proportional to the number of nodes in a cell. Since the assumption is that the sensor nodes are uniformly distributed, the number of nodes in a cell is proportional to the volume of the cell. So in general, the ratio of network lifetime in different models is essentially the ratio of volume of a cell under those models. Then network lifetime of CB model is $\dfrac{17\sqrt{17}}{96\sqrt{3}} =$ 42.154% of that of *TO* model. It is $\dfrac{17\sqrt{17}}{108} = 64.9\%$, for *Alt-CB*, $\dfrac{17\sqrt{17}}{28\sqrt{14}} = 66.9\%$ for *HP* model, $\dfrac{3\sqrt{3}}{4\sqrt{2}} = 91.86\%$ for *Alt-HP* and $\dfrac{17\sqrt{17}}{128} = 54.76\%$ for *RD* model (See Figure 18).

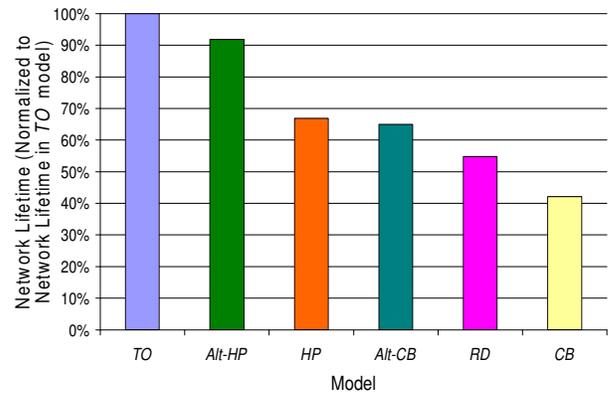

Figure 18: Network lifetime in various models

### 4.1.4 Closeness to Optimality

Clearly, our GAF like approach of dividing a network into cells and keeping one node active in each cell is suboptimal. In this subsection, we estimate how close to optimality our scheme is in terms of number of nodes kept active at any time. According to our analysis, in 2-D case, the number of nodes required by 2D-GAF is 4 times of that the optimal number and in 3-D case this value is 8 times. Although we cannot improve this scenario for 1-coverage, GAF-like approach may require significantly fewer nodes for *k*-coverage.

*2D GAF:*

Let us first explore how to improve 2D-GAF for k-coverage. For 1-coverage, we have to keep one node active in a hexagonal cell with $r = \dfrac{r_s}{2}$, where $r_s$ is sensing range of each sensor. For k-coverage, we can set the radius of each cell $r = \dfrac{r_s}{2\sqrt{\lceil k/4 \rceil}}$ and still keep one node active in each cell. Then the area of each cell is $= \dfrac{3\sqrt{3}}{2}r^2 = \dfrac{3\sqrt{3}}{2}\dfrac{r_s^2}{4\lceil k/4 \rceil}$. So total number of cells within $r_s$ distance of any point is

$$\dfrac{\pi r_s^2}{\dfrac{3\sqrt{3}}{2}\dfrac{r_s^2}{4\lceil k/4 \rceil}} = \dfrac{8\pi \lceil k/4 \rceil}{3\sqrt{3}} \geq \dfrac{2\pi k}{3\sqrt{3}} = 1.209199576k$$

Note that $2\pi/3\sqrt{3}$ is the ratio of area of a circle and a hexagon of equal radius.

For the next step, we need the help of the following theorem.

**Theorem 4.1:** The sum of two independent *Poisson* random variables is *Poisson* as well, with parameter equal to the sum of the two individual parameters.

**Proof:** See appendix.

According to the above theorem, if the two areas have the same node density $\rho_1 = \rho_2 = \rho$, then

$$P(K_1 + K_2 = k) = e^{-\rho(a_1+a_2)} \dfrac{[\rho(a_1+a_2)]^k}{k!}$$; i.e., one

can just simply "combine" non-overlapping areas for the purpose of calculating the parameter of the *Poisson* distribution.

Since we have active node density $\rho = \dfrac{1}{\dfrac{3\sqrt{3}}{2}\dfrac{r_s^2}{4\lceil k/4 \rceil}}$ node per unit area, or in other word, $\rho = 1$ node per cell. Within $r_s$ distance of any point, the number of active nodes is a *Poisson* random variable $K$ with parameter $\lambda_k = \dfrac{8\pi \lceil k/4 \rceil}{3\sqrt{3}}$.

We want $P(K \geq k)$ to be high and fairly close to 1. Now,

$$P(K \geq k) = 1 - P(K < k)$$
$$= 1 - \sum_{i=0}^{k-1} P(K = i) = 1 - \sum_{i=0}^{k-1} e^{-\lambda_k}\dfrac{\lambda_k^i}{i!}$$
$$= 1 - \sum_{i=0}^{k-1} e^{-\left(\dfrac{8\pi\lceil k/4\rceil}{3\sqrt{3}}\right)} \dfrac{\left[\dfrac{8\pi\lceil k/4\rceil}{3\sqrt{3}}\right]^i}{i!}$$

i.e., $P(K \geq k) = 1 - e^{-\left(\dfrac{8\pi\lceil k/4\rceil}{3\sqrt{3}}\right)} \sum_{i=0}^{k-1} \dfrac{[8\pi\lceil k/4 \rceil]^i}{[3\sqrt{3}]^i i!}$

| K | $\lambda_k$ | P(K>=k) | Number of nodes vs Optimal= $\dfrac{4\lceil k/4 \rceil}{k}$ |
|---|---|---|---|
| 1 | 4.8367983 | 1 | 400% |
| 2 | 4.8367983 | 0.9616325 | 200% |
| 3 | 4.8367983 | 0.8688446 | 133% |
| 4 | 4.8367983 | 0.7192460 | 100% |
| 5 | 9.6735966 | 0.9639949 | 160% |

*3D GAF:*

For 1-coverage, we have to keep one node active in a truncated octahedron cell with $r = \dfrac{r_s}{2}$, where $r_s$ is sensing range of each sensor. For k-coverage, we can set the radius of each cell $r = \dfrac{r_s}{2\sqrt[3]{\lceil k/8 \rceil}}$ and still keep one node active in each cell. Then the volume of each cell is $= \dfrac{32}{5\sqrt{5}}r^3 = \dfrac{32}{5\sqrt{5}}\dfrac{r_s^3}{8\lceil k/8 \rceil}$. So total number of cells within $r_s$ distance of any point is

$$\dfrac{\dfrac{4}{3}\pi r_s^3}{\dfrac{32}{5\sqrt{5}}\dfrac{r_s^3}{8\lceil k/8 \rceil}} = \dfrac{5\sqrt{5}\pi \lceil k/8 \rceil}{3}$$
$$\geq \dfrac{5\sqrt{5}\pi k}{24} = 1.4635030689k$$

Note that $5\sqrt{5}\pi/24$ is the ratio of volume a sphere and a truncated octahedron of equal radius. Since we have active node density $\rho = \dfrac{1}{\dfrac{32}{5\sqrt{5}} \dfrac{r_s^3}{8\lceil k/8 \rceil}}$ node per unit volume, or in other word, $\rho = 1$ node per cell. Within $r_s$ distance of any point, the number of active nodes is a *Poisson* random variable $K$ with parameter $\lambda_k = \dfrac{5\sqrt{5}\pi \lceil k/8 \rceil}{3}$.

We want to $P(K \geq k)$ to be high and fairly close to 1. Now,

$$P(K \geq k) = 1 - P(K < k)$$
$$= 1 - \sum_{i=0}^{k-1} P(K = i) = 1 - \sum_{i=0}^{k-1} e^{-\lambda_k} \frac{\lambda_k^i}{i!}$$
$$= 1 - \sum_{i=0}^{k-1} e^{-\left(\frac{5\sqrt{5}\pi \lceil k/8 \rceil}{3}\right)} \frac{\left[\frac{5\sqrt{5}\pi \lceil k/8 \rceil}{3}\right]^i}{i!}$$
$$= 1 - e^{-\left(\frac{5\sqrt{5}\pi \lceil k/8 \rceil}{3}\right)} \sum_{i=0}^{k-1} \frac{\left[5\sqrt{5}\pi \lceil k/8 \rceil\right]^i}{3^i i!}$$

| k | $\lambda_k$ | P(K>=k) | Number of nodes vs Optimal= $\dfrac{8\lceil k/8 \rceil}{k}$ |
|---|---|---|---|
| 1 | 11.70802455 | 1 | 800% |
| 2 | 11.70802455 | 0.9999 | 400% |
| 3 | 11.70802455 | 0.9994 | 233% |
| 4 | 11.70802455 | 0.9971 | 200% |

So our 3D-GAF scheme achieves 4-coverage with probability 0.9971 with twice the optimal number of nodes.

## 5. DISCUSSIONS AND FUTURE WORKS

Following discussions are applicable to both active nodes in nonhierarchical networks and backbone nodes in hierarchical networks, and so we use the general term node to refer to both types of nodes in the two network architectures described in this paper. Nodes can use their cell id as their address. A greedy geographic routing scheme can work here as follows: source node writes its cell id and destination node's cell id in the packet. Suppose that the source cell id is $(u_s, v_s, w_s)$ and the destination cell id is $(u_d, v_d, w_d)$. Then the source sends this packet to a neighbor with cell id $(u_i, v_i, w_i)$ such that $(u_d - u_i)^2 + (v_d - v_i)^2 + (w_d - w_i)^2 < (u_d - u_s)^2 + (v_d - v_s)^2 + (w_d - w_s)^2$. Then the node with cell id $(u_i, v_i, w_i)$ sends this packet to a neighbor with cell id $(u_j, v_j, w_j)$ such that $(u_d - u_j)^2 + (v_d - v_j)^2 + (w_d - w_j)^2 < (u_d - u_i)^2 + (v_d - v_i)^2 + (w_d - w_i)^2$. If more than one neighbor satisfies above criteria (most often which is actually the case), then the least loaded node, the node with the highest energy or just randomly one of them can be chosen. When the shape of each cell is truncated octahedron, each cell has 14 neighboring cells. The neighboring cells of a cell having cell id $(u_1, v_1, w_1)$ have the following ids: $(u_1+1, v_1, w_1)$, $(u_1-1, v_1, w_1)$; $(u_1, v_1+1, w_1)$, $(u_1, v_1-1, w_1)$; $(u_1-1, v_1-1, w_1+2)$, $(u_1+1, v_1+1, w_1-2)$; $(u_1, v_1, w_1+1)$, $(u_1, v_1, w_1-1)$; $(u_1-1, v_1, w_1+1)$, $(u_1+1, v_1, w_1-1)$; $(u_1, v_1-1, w_1+1)$, $(u_1, v_1+1, w_1-1)$; $(u_1-1, v_1-1, w_1+1)$, $(u_1+1, v_1+1, w_1-1)$. So it requires a small constant number of arithmetic operations to choose the optimal neighboring node to forward a packet. Above simple approach works well when all nodes are always connected with all of their neighboring nodes. However, this greedy scheme might not work in all possible scenarios. In the presence of obstacle, there is a possibility that the packet reaches a dead end where there is no neighboring node that satisfies the criteria mentioned above and the packet is yet to reach the destination. Routing in such cases for 3D network has been investigated in [9][11].

Since this paper relies heavily on localization in 3D underwater environment, one potential future work can be finding a robust mechanism for localization in 3D underwater environments.

## 6. CONCLUSION

In this paper, we provide two different network architectures for 3D underwater wireless sensor networks. The first architecture is a hierarchical network consisting of backbone nodes responsible for creating and maintaining the network. We provide a placement strategy that minimizes the number of backbone nodes needed while keeping the network fully functional. We also discuss frequency reuse for this 3D network and energy efficiency issues of our proposed schemes. The mobile sensors that communicate directly with the nearest backbone node are free to move as long as there are sufficient mobile sensors everywhere for sensing purpose. So a dense, uniform and random deployment of mobile sensors is sufficient. In the second architecture, we assume there backbone nodes are not available to create and maintain the network. So sensing as well as communicating with the sink is the responsibility of the sensors themselves. We partition the 3D space into identical cells and keep one node active in each cell inspired by GAF in 2D network. We analyze six most likely partitioning schemes in 3D and find that partitioning the 3D space into truncated octahedron shaped cells is the best approach. In this case, full coverage can be achieved if the sensing range is at least 0.542326 times the transmission radius. We also provide a distributed algorithm that allows a sensor node to determine its cell id using a few simple local arithmetic operations provided that the location information is available. We also provide closeness to optimality of our proposed scheme.

## 7. REFERENCES

[1] I. F. Akyildiz, D. Pompili and T. Melodia, "Underwater Acoustic Sensor Networks: Research Challenges", *Ad Hoc Networks Journal*, *(Elsevier)*, March 2005.

[2] S. M. N. Alam and Z. Haas, "Coverage and Connectivity in Three-Dimensional Networks", In *Proc of ACM MobiCom*, 2006.

[3] S. M. N. Alam and Z. Haas, "Coverage and Connectivity in Three-Dimensional Underwater Sensor Networks", *Wireless Communication and Mobile Computing*, 8: 995-1009, October, 2008.

[4] Aristotle, *On the Heaven*, Vol. 3, Chapter 8, 350BC.

[5] E. S. Barnes and N. J. A. Sloane, "The Optimal Lattice Quantizer in Three Dimensions", *SIAM J. Algebraic Discrete Methods* 4, 30-41, 1983.

[6] S. Basagni, A. Carosi and C. Petrioli, "Sensor DMAC: Dynamic Topology Control for Wireless Sensor Networks," In *Proc of IEEE VTC,* September 2004.

[7] B. Chen, K. Jamieson, H. Balakrishnan, and R. Morris."Span: an energy-efficient coordination algorithm for topology maintenance in ad hoc wireless networks". *Wireless Networks*,8(5), 2002.

[8] W. Cheng, A.Y. Teymorian, L. Ma, X. Cheng, X. Lu and Z. Lu, "Underwater Localization in Sparse 3D Acoustic Sensor Networks", pp.236-240, *INFOCOM 2008*.

[9] S. Durocher, D. Kirkpatrick and L. Narayanan, "On Routing with Guaranteed Delivery in Three-Dimensional Ad Hoc Wireless Networks", *ICDCN 2008*.

[10] A. Ellatifi, "Tiling and 3D Frequency Planning", pp. 291-296, *PIMRC 1994*.

[11] R. Flury and R. Wattenhofer, "Randomized 3D Geographic Routing", *INFOCOM 2008*.

[12] M. Gardner, The Sixth Book of Mathematical Games from Scientific American, Chicago, IL: University of Chicago Press, 1984.

[13] J. Heidemann, W. Ye, J. Wills, A. Syed, and Y. Li. "Research Challenges and Applications for Underwater Sensor Networking"*, Wireless Communications and Networking Conference,* 2006.

[14] D. Hilbert and S. Cohn-Vossen. *Geometry and the Imagination.* New York: Chelsea, 1999.

[15] N. W. Johnson. *Uniform Polytopes.* Cambridge, England: Cambridge University Press, 2000.

[16] J. Kong, J. Cui, D. Wu, M. Gerla, "Building Underwater Ad-hoc Networks and Sensor Networks for Large Scale Real-time Aquatic Applications", *IEEE MILCOM'05*.

[17] M. Křížek, "Superconvergence phenomena on three-dimensional meshes", *International Journal*

of *Numerical Analysis and Modeling*. Vol. 2, No. 1, 2005, pp. 43-56.

[18] N. Lynch, *Distributed Algorithms*, Morgan Kaufmann Publishers, Wonderland, 1996.

[19] D. Pompili and T. Melodia, "Three-dimensional Routing in Underwater Acoustic Sensor Networks," in *Proc. of ACM International Workshop on Performance Evaluation of Wireless Ad Hoc, Sensor, and Ubiquitous Networks (PE-WASUN)*, Montreal, Canada, October 2005

[20] D. Pompili, T. Melodia and I. F. Akyildiz, "Three-dimensional and two-dimensional deployment analysis for underwater acoustic sensor networks", *Ad Hoc Networks*, 7(4), 778-790, June 2009.

[21] R. Lei, L. Wenyu and G. Peng, "A coverage algorithm for three-dimensional large-scale sensor network", *International Symposium on Intelligent Signal Processing and Communication Systems (ISPACS)*, pp 420-423, 2007.

[22] H. Steinhaus. *Mathematical Snapshots*, 3rd edition, Oxford University Press, 1969.

[23] M. Stojanovic, "Design and Capacity Analysis of Cellular Type Underwater Acoustic Networks," *IEEE Journal of Oceanic Engineering*, 33(2), pp.171-181, April 2008.

[24] M. Stojanovic, "Frequency Reuse Underwater: Capacity of an Acoustic Cellular Network," in *Proc. Second ACM International Workshop on Underwater Networks (WUWNeT'07)*, Montreal, Canada, September 2007.

[25] W. Thomson (Lord Kelvin), "On the division of space with minimum partition area". *Philosophical Magazine*, 24 (1887)503-514.

[26] D. Weaire and R. Phelan. "A Counter-Example to Kelvin's Conjecture on Minimal Surfaces". *Philosophical Magazine Letters*, 69, 107-110, 1994

[27] D. Wells. The Penguin Dictionary of Curious and Interesting Geometry, Penguin, UK, 1991.

[28] H. Weyl. *Symmetry.* Princeton, NJ: Princeton University Press, 1952.

[29] Y. Xu, J. Heideman and D. Estrin. "Geography-informed energy conservation in ad hoc routing". In *Proc. of the 7th ACM MobiCom*, July, 2001.

[30] F. Ye, G. Zhong, S. Lu, L. Zhang. "PEAS: A Robust Energy Conserving Protocol for Long-lived Sensor Networks", *ICDCS '03*, Rhode Island, May 2003.

[31] H. Zhang and J. C. Hou, "Maintaining sensing coverage and connectivity in large sensor networks," *Wireless Ad Hoc and Sensor Networks: An International Journal*, 1, 89-124, 2005.

**Appendix:**

**Theorem 4.1:** The sum of two independent *Poisson* random variables is *Poisson* as well, with parameter equal to the sum of the two individual parameters.

*Proof:* Assume that we have two areas, $A_1 [m^2]$ and $A_2 [m^2]$, where in each area nodes are randomly distributed based on 2D *Poisson* distribution with parameters $\rho_1$ [nodes/m$^2$] and $\rho_2$ [nodes/m$^2$], respectively. Within the areas $A_1$ and $A_2$, there are sub-areas, $a_1 [m^2]$ and $a_2 [m^2]$, respectively, which are chosen independently one from the other. Consequently, the expected numbers of nodes in the two sub-area $a_1$ and $a_2$ are: $a_1 \rho_1$ and $a_2 \rho_2$, respectively, and the number of nodes within each sub-area is also *Poisson* with the parameters $\lambda_1 = a_1 \rho_1$ and $\lambda_2 = a_2 \rho_2$, respectively (this can be easily shown).

Let's label as $K_1$ and $K_2$ as the random variable indicating the number of nodes in the areas $a_1$ and $a_2$, respectively. We can write that the probabilities of finding $k$ nodes in area $a_1$ and area $a_2$ are, respectively:

$$P(K_1 = k) = e^{-\lambda_1} \frac{\lambda_1^k}{k!} \text{ and } P(K_2 = k) = e^{-\lambda_2} \frac{\lambda_1^k}{k!}.$$

We are trying to show that $P(K_1 + K_2 = k) = e^{-(\lambda_1 + \lambda_2)} \frac{(\lambda_1 + \lambda_2)^k}{k!}$; i.e., the probability that the total number of nodes in both sub-areas, $K = k$, is also *Poisson* with the parameter equal to the expected total number of nodes in the two area, $\lambda = \lambda_1 + \lambda_2 = a_1\rho_1 + a_2\rho_2$.

$$P(K_1 + K_2 = k)$$
$$= \sum_{i=0}^{k} P(K_1 + K_2 = k | K_2 = i) \times P(K_2 = i)$$
$$= \sum_{i=0}^{k} P(K_1 = k-i | K_2 = i) \times P(K_2 = i)$$
$$= \sum_{i=0}^{k} P(K_1 = k-i) \times P(K_2 = i)$$
$$= \sum_{i=0}^{k} \left[ \left( e^{-l_1} \frac{l_1^{k-i}}{(k-i)!} \right) \times \left( e^{-l_2} \frac{l_2^{i}}{i!} \right) \right]$$
$$= e^{-(\lambda_1+\lambda_2)} \sum_{i=0}^{k} \left[ \left( \frac{1}{(k-i)!i!} \right) \lambda_1^{k-i} \lambda_2^{i} \right]$$
$$= \frac{e^{-(\lambda_1+\lambda_2)}}{k!} \sum_{i=0}^{k} \left[ \left( \frac{k!}{(k-i)!i!} \right) \lambda_1^{k-i} \lambda_2^{i} \right]$$
$$= \frac{e^{-(\lambda_1+\lambda_2)}}{k!} \sum_{i=0}^{k} \left[ \binom{k}{i} \lambda_1^{k-i} \lambda_2^{i} \right]$$
$$= \frac{e^{-(\lambda_1+\lambda_2)}}{k!} (\lambda_1 + \lambda_2)^k$$
$$= \frac{e^{-(a_1\rho_1+a_2\rho_2)}}{k!} (a_1\rho_1 + a_2\rho_2)^k$$

Thus, if we have two independent *Poisson* random variables, the sum of the two variables is *Poisson* as well, with parameter equal to the sum of the two individual parameters.

Note #1: in the proof above, we used the fact that the two random variables are independent by using the fact:

$P(K_1 = k-i | K_2 = i) = P(K_1 = k-i)$.

Note #2: By repeating the process *n*-1 times, we can prove that the sum *n* independent *Poisson* random variable is *Poisson* as well with parameter equal to the sum of the *n* parameters of the individual random variables.

In particular, for our application of the sum of the number of nodes in two (independently) selected sub-areas is *Poisson* with parameter equal to the sum of the expected number of nodes in each individual area. Note that this is a valid statement even if the two sub-areas are in the same area, as long as there is no overlap between the two sub-areas.

If the two areas have the same node density $\rho_1 = \rho_2 = \rho$, then

$P(K_1 + K_2 = k) = e^{-\rho(a_1+a_2)} \frac{[\rho(a_1+a_2)]^k}{k!}$; i.e., one can just simply "combine" the areas for the purpose of calculating the parameter of the *Poisson* distribution. Note that this would be the case when the two sub-areas are both within the same area (and are, of course, non-overlapping).